\documentstyle[multicol,epsfig,prc,aps]{revtex}  
\begin{document}  
%
%
%
%
\def\mys#1{\Sigma_{ } { }_{\bf #1}}  
\def\ro{\rho_{\rm o}}  
\def\reff{$R_{eff}$}   
\def\lp{\left(}   
\def\rp{\right)}    
\def\bc{\begin{center}}  
\def\ec{\end{center}}  
\def\vp{\vspace*{0.1cm}}  
\def\vm{\vspace*{-0.2cm}}  
\def\CF{{\it CF}\,\,}   
\def\CFful{{Cooper-Frye}\,\,}  
\def\CO{{\it CO}\,\,}   
\def\COful{{\it cut-off}\,\,}   
%
%
%
%
  
\title{  
Boundary Conditions of the Hydro-Cascade Model and 
Relativistic Kinetic Equations for Finite Domains}

\author{\bf K.A. Bugaev} 
\address{  
Bogolyubov Institute for Theoretical Physics,  
Kiev, Ukraine\\  
Lawrence Berkeley National Laboratory, Berkeley, California, USA 
}  

\date{\today}  

\maketitle

\noindent 
\begin{abstract}
 A detailed analysis of 
the coupled  relativistic kinetic equations for two 
domains separated by a  hypersurface  having  both space- and time-like 
parts is presented. 
Integrating the derived set of transport  equations, we obtain 
the correct system of the hydro+cascade  equations to model the relativistic 
nuclear collision process.
Remarkably,  the conservation laws on the
boundary between domains conserve separately  both  the incoming  
and outgoing components of energy, momentum and baryonic charge. Thus,
the  relativistic kinetic theory generates twice the number of  conservation laws 
compared to 
traditional hydrodynamics.
Our analysis shows that these boundary conditions between domains,
the {\it three flux discontinuity},
can be  satisfied only by a special superposition of two {\it cut-off} distribution functions for
the ``out'' domain.
All these results are applied to the case of the phase transition between 
quark gluon plasma and hadronic matter. The possible  consequences for an improved
hydro+cascade description of the relativistic nuclear collisions are  discussed.
The unique properties   of the {\it three flux discontinuity} and  their effect on the 
space-time evolution of the transverse expansion are also analyzed. 
The possible modifications of both transversal radii  from   pion correlations 
generated by a correct hydro+cascade approach
are discussed.  
\end{abstract} 
 
\vspace*{0.2cm} 

\noindent
\hspace*{1.9cm}{\it PACS numbers:} 25.75.Ld

\vspace*{0.2cm} 
  
\noindent 
\hspace*{1.9cm}\begin{minipage}[t]{14.2cm} 
{\bf Key words:} 
kinetic equations with source terms, hydro+cascade  equations, conservation laws,
three flux discontinuity 
\end{minipage}

\begin{multicols}{2}  
  

\section{Introduction}  
  
The modern history of relativistic hydrodynamics 
started more than fifty years ago when  L. D. Landau  
suggested \cite{LANDAU:53}  its use  to describe the expansion 
of the strongly interacting matter that is 
formed in  high energy hadronic collisions.   
Since that time there arose a fundamental problem 
of relativistic hydrodynamics known as the  freeze-out problem.   
In other words,  one has to know  
how to stop solving the hydrodynamical equations and convert  
the matter into free streaming particles.  
There were   several ways suggested  to handle it,   
but  only   recently  a  new approach 
to solve the freeze-out problem  in relativistic hydrodynamics 
has been  invented  by   Bass and Dumitru (BD model)  \cite{BD:00}
and further developed  by 
Teaney, Lauret and Shuryak (TLS model)  \cite{SHUR:01}. 
These  hydro + cascade   models assume that  the nucleus-nucleus collisions   
proceed in three stages: hydrodynamic   
expansion (hydro) of the quark gluon plasma (QGP), phase transition from the QGP to  
the hadron gas (HG) and the stage of hadronic  
rescattering and resonance decays (cascade). The switch from hydro to  
cascade modeling takes place   
at the boundary between the mixed  and hadronic phases.   
The spectrum of hadrons  
leaving this hypersurface of the QGP--HG transition is taken as input for the  
cascade.  
  
This  approach  incorporates  the best features of  both the hydrodynamical and cascade 
descriptions.  It  allows for, on one hand, the calculation of  the phase transition 
between  the quark gluon plasma  and  hadron gas using hydrodynamics  
and, on the other hand,  the freeze-out of hadron spectra   
using  the cascade description. 
This  approach 
allows  one to  overcome  the  usual 
difficulty of  transport models  in modeling    phase transition phenomenon. 
For this reason, this approach has been rather successful in explaining  
a variety of collective phenomena that has been observed
at the CERN  Super Proton Collider (SPS)  and  
Brookhaven Relativistic Heavy Ion Collider (RHIC) energies.  
However, both the BD and TLS models face some   
fundamental  difficulties    
which cannot be ignored {(see a detailed discussion in \cite{BUG:02}).}  
Thus, within the BD approach   
the initial distribution for the cascade is found using  the \CFful formula \cite{COOP:74},  
which takes into account particles with all possible velocities,   
whereas in the TLS model the initial cascade distribution is given by the \COful formula  
\cite{BUG:96,BUG:99a},  
which accounts for only those particles that can leave the phase boundary.   
As shown in Ref. \cite{BUG:02}  the \CFful formula leads to  causal and  
mathematical problems in the present version of the  BD model because the  
QGP--HG phase boundary inevitably has time-like parts.  
On the other hand, the TLS model  does not conserve  
energy, momentum and number of charges and this, as will be demonstrated later,  
is due to the fact that the equations of motion used in \cite{SHUR:01}  
are incomplete and, hence, should be modified. 
 
These difficulties are likely in part responsible for the fact 
that the existing  hydro+cascade  models, like  the more simplified ones,  
fail to  explain  the {\it HBT puzzle} \cite{GYULASSY:01}, i.e. 
the fact that  
the experimental HBT radii at RHIC   are  very similar to those
found at SPS, even though  the centre of mass energy is larger  by an order of 
magnitude. Therefore, it turns out that  the  hydro+cascade approach 
successfully {\it parameterizes}  the one-particle momentum spectra and their 
moments, but does not {\it  describe} the space-time picture of the  
nuclear collision  as probed by  two-particle interferometry.  
 
The main  difficulty of the hydro + cascade  approach looks  similar to  
the traditional problem of 
freeze-out  in relativistic hydrodynamics \cite{BUG:96,BUG:99a}. 
In both cases the  domains (subsystems) have time-like boundaries 
through which the exchange of particles  occurs  and this fact should be taken into account.  
In relativistic hydrodynamics this problem was solved by the  
constraints which appear  on the freeze-out  
hypersurface and provide  the global energy-momentum and  
charge conservation \cite{BUG:96,BUG:99a,BUG:99b}. 
{ A generalization of the usual} 
Boltzmann equation which accounts  
for  the exchange of particles 
on the time-like  boundary between domains in the relativistic kinetic theory 
was given recently  in Ref.  \cite{BUG:02}. 
It was  shown  that
the kinetic equations  
describing the exchange of particles   
on the time-like  boundary between subsystems   
should necessarily contain  the $\delta$-like  source terms. 
{ From these kinetic equations} 
the correct system of   hydro+cascade  equations to  model the relativistic 
nuclear collision process was derived 
without specifying the properties of 
the separating hypersurface. 
{ However, both an  explicit switch off} criterion 
from the hydro equation to the cascade one  
and the boundary conditions between them   
were not considered in \cite{BUG:02}. 
The  present work  is devoted  
to the analysis of the boundary  
conditions for the system of   hydro+cascade  equations. 
This is 
necessary to formulate the numerical algorithm 
for solving the hydro+cascade  equations.

The paper is organized as follows. 
In Sect.~2
a brief derivation of the set of    kinetic equations 
is given and  source terms are obtained. 
In Sect.~3  the analog of the collision integrals 
is  discussed and a fully covariant formulation of the system of coupled 
kinetic equations is found.   The relation between the system obtained and 
the relativistic Boltzmann equation is also considered. 
The correct equations of motion for the  hydro + cascade  approach 
and their boundary conditions 
are analyzed in Sect.~4. 
There it is  also shown that the existence of  strong discontinuities across the space-like boundary,
the time-like shocks,  is in contradiction with  the basic assumptions of  a transport approach.
The solutions of   boundary conditions between  the hydro and cascade domains for a single degree of freedom
and for many degrees of freedom are discussed in Sect. 5 and 6, respectively.
The conclusions are given in Sect.~7.

\section{Drift Term for Semi-Infinite Domain} 
 
\vm

Let us consider  two semi-infinite domains, ``in'' and ``out'',  
separated by the hypersurface $\Sigma^*$  
which, for the purpose of presenting the idea, we assume to be 
given in (3+1) dimensions   
by a single valued function  
$t = t^*(\bar{x}) = x_0^*(\bar{x})$.  
{ The latter}  
is assumed to be a unique solution of the equation 
${\cal F}^{*}(t, \bar{x}) = 0$ ({\it a switch off criterion}) 
which has  a positive time derivative 
$\partial_0 {\cal F}^{*}(t^*, \bar{x}) > 0$ on the hypersurface $\Sigma^*$. 
The distribution function $\phi_{in}(x,p)$  for $t \le t^*(\bar{x})$    
is assumed to belong  
to the ``in'' domain,  
whereas $\phi_{out}(x,p)$ denotes the distribution function of the ``out'' domain   
for $t \ge t^*(\bar{x})$(see Fig.1).  
In this work it is assumed that the initial conditions for   
$\phi_{in}(x,p)$ are given, whereas   
on $\Sigma^*$ the  function $\phi_{out}(x,p)$ is allowed  
to differ from $\phi_{in}(x,p)$ and   
this will modify the kinetic equations for both functions.  
For simplicity we  consider a classical gas of point-like Boltzmann particles.  

\vspace*{0.3cm}

\begin{figure}
\hspace*{0.0cm}\mbox{ \psfig{file=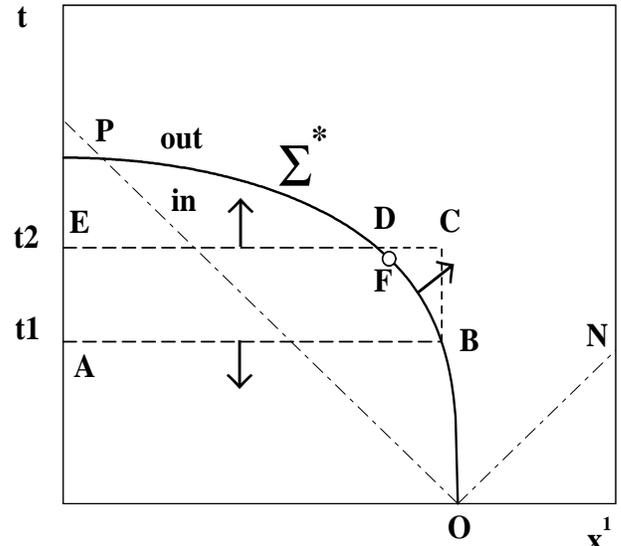,width=80mm,height=75mm} }
\caption{
Schematic two dimensional picture
of the boundary hypersurface $\Sigma^*$ (solid curve).
Arrows show the external normal
vectors. The light cone $NOP$ is shown by the dash-dotted line.
The point $F$ divides $\Sigma^*$ into the time-like  ($OF$) and space-like ($FP$) parts.
}
\label{fig1}
\end{figure}

\vspace*{-0.2cm}

Similar to Ref. \cite{GROOT} we  derive the kinetic equations  
for $\phi_{in}(x,p)$ and $\phi_{out}(x,p)$ from the requirement of    
particle number conservation.   
Therefore, the particles leaving one domain and crossing the  hypersurface $\Sigma^*$   
should be subtracted from the corresponding distribution function and  
added to the other.  
Now  consider the closed hypersurface  of the ``in'' domain, $\Delta x^3$   
(shown as the contour $ABDE$ in Fig.1),    
which consists of  
two semi-planes $\sigma_{t1}$ and $\sigma_{t2}$ of constant time $t1$ and $t2$, respectively,  
that are connected from $t1$ to $t2 > t1$ by  
the arc $BD$  of the boundary $\Sigma^*(t1,t2)$    
in Fig.1.  
The original number of particles on the hypersurface $\sigma_{t1}$ is given by   
the standard expression \cite{GROOT}  
 
\vm 
\begin{equation}\label{one}  
N_1 = - \int\limits_{\sigma_{t1}} d \Sigma_\mu \frac{d^3 p}{p^0}~ p^\mu~ \phi_{in}(x,p)\,,   
\end{equation}  
\vm

\noindent  
where $d \Sigma_\mu$ is  the external normal vector to $\sigma_{t1}$ and, hence,   
the product $p^\mu d \Sigma_\mu \le 0$ is non-positive.   
It is clear that these particles can   
cross either hypersurface $\sigma_{t2}$ or $\Sigma^*(t1,t2)$.  
The corresponding numbers of particles are as follows  

\vm  
\begin{eqnarray}\label{two}  
N_2  \hspace*{0.0cm}=   \hspace*{-0.1cm}\int\limits_{\sigma_{t2}} \hspace*{-0.1cm}&& d \Sigma_\mu \frac{d^3 p}{p^0}~ p^\mu~ \phi_{in}(x,p)\,,\\   
\label{three}  
N_{loss}^*  = \hspace*{-0.2cm}\int\limits_{\Sigma^*(t1,t2)}\hspace*{-0.4cm}&&  d\Sigma_\mu   
\frac{d^3 p}{p^0}~ p^\mu~ \phi_{in}(x,p) ~\Theta(p^\nu d\Sigma_\nu) \,.  
\end{eqnarray}  
\vm  
\vm  
  
\noindent  
The $\Theta$-function in the {\it loss} term (\ref{three}) is very important  
because it accounts for the particles leaving the ``in'' domain  (see also  
discussion in \cite{BUG:96,BUG:99b}).   
For the space-like parts of the hypersurface  $\Sigma^*(t1,t2)$ which are   
defined by negative  sign $ds^2 < 0$ of   
the squared line  element, $ds^2 = dt^*(\bar{x})^2 - d\bar{x}^2 $,   
the product $p^\nu d\Sigma_\nu > 0$ is always positive and, therefore,   
particles with all possible  momenta can leave the ``in'' domain through  
the $\Sigma^*(t1,t2)$.  
For the time-like parts of $\Sigma^*(t1,t2)$ (with sign $ds^2 > 0$)   
the product $p^\nu d\Sigma_\nu $ can have either sign, and the $\Theta$-function  
{\it cuts off} those particles which return to the ``in'' domain.  
  
Similar one has to consider the particles coming to the  ``in'' domain   
from outside. This is possible through the time-like parts of the hypersurface $\Sigma^*(t1,t2)$,   
if the  particle momentum satisfies  the inequality $ - p^\nu d\Sigma_\nu > 0$.   
In terms of the external normal $d \Sigma_\mu$ with respect to the ``in'' domain   
(this normal vector is   
shown as an arrow on the arc $BD$  in Fig.1 and   
will be used hereafter for all integrals over the  hypersurface $\Sigma^*(t1,t2)$)   
the number of gained particles    
  
\vm   
\begin{equation}\label{four}  
N_{gain}^*  = - \int\limits_{\Sigma^*(t1,t2)}\hspace*{-0.4cm}  d\Sigma_\mu  
\frac{d^3 p}{p^0}~ p^\mu~ \phi_{out}(x,p) ~\Theta(-p^\nu d\Sigma_\nu) \,  
\end{equation}  
\vm  
\vm  
   
\noindent  
is, evidently, non-negative.    
Since the total number of particles is conserved, i.e.   
$N_2 = N_1 - N_{loss}^* + N_{gain}^*$, one can use the Gauss theorem  
to rewrite the obtained integral over the closed hypersurface $\Delta x^3$   
as an  integral over the  $4$-volume $\Delta x^4 $   
(area inside  the contour $ABDE$ in Fig.1)   
surrounded  by $\Delta x^3$  

\vm  
\begin{eqnarray}  
\int\limits_{\Delta x^4} \hspace*{-0.1cm}  d^4 x  
\frac{d^3 p}{p^0}~ p^\mu  ~{\partial}_\mu ~ \phi_{in}(x,p) = \int\limits_{\Sigma^*(t1,t2)}\hspace*{-0.4cm}  d\Sigma_\mu \frac{d^3 p}{p^0}~ p^\mu \times  && \nonumber \\   
\label{five}   
\vm  
%
[\phi_{in}(x,p) - \phi_{out}(x,p) ] \Theta(-p^\nu d\Sigma_\nu) \,.&&  
\end{eqnarray}  
\vm  
\vm  
   
\noindent  
Note that in contrast to the usual case \cite{GROOT}, i.e. in the absence of 
a boundary $\Sigma^*$,     
the right-hand side (rhs) of Eq. (\ref{five})  does not vanish identically.  
  
The rhs of Eq. (\ref{five}) can be transformed further to  
a  $4$-volume integral in the following  
sequence of steps. First we express the integration element $d\Sigma_\mu$  
via the normal vector $n^*_\mu$ as follows $(dx^j > 0,$ for $ j =1,2,3)$   

\vm  
\begin{equation}\label{six}  
 d\Sigma_\mu = n^*_\mu dx^1 dx^2 dx^3;   
 \quad  n^*_\mu \equiv \delta_{\mu 0} - \frac{ \partial t^*(\bar{x}) }{\partial x^\mu} (1 - \delta_{\mu 0} )\,,   
\end{equation}  
\vm  
\vm  
   
\noindent  
where $\delta_{\mu \nu}$ denotes the Kronecker symbol.   
Then, using the identity  $\int\limits_{t1}^{t2} dt\, \delta (t - t3) = 1$    
for the Dirac $\delta$-function  
with  
$t1 \le t3 \le t2$, we rewrite the rhs integral in (\ref{five}) as   

\vm  
\begin{equation}\label{seven}   
\int\limits_{\Sigma^*(t1,t2)}\hspace*{-0.4cm}  d\Sigma_\mu \cdots \equiv   
\int\limits_{V^4_\Sigma} d^4 x~\delta (t - t^*(\bar{x}) )~ n^*_\mu \cdots\,,   
\end{equation}  
\vm  
\vm  
   
\noindent  
where the $4$-dimensional volume $V^4_\Sigma$ is  
a direct product of the $3$- and $1$-dimensional  
volumes $\Sigma^*(t1,t2)$ and $(t2-t1)$, respectively.   
Evidently, the Dirac $\delta$-function allows us to extend integration in (\ref{seven}) to the  
unified $4$-volume  $V^4_U = \Delta x^4 \cup V^4_\Sigma$ of $\Delta x^4$ and $V^4_\Sigma$  
(the volume  $V^4_U $ is shown as the area $ABCE$ in Fig.1).  
Finally, with the help of notations  

\begin{equation}\label{eight}  
\Theta_{out} \equiv \Theta (t - t^*(\bar{x}) ); \quad \Theta_{in} \equiv 1- \Theta_{out}   
\end{equation}  
it is possible to extend the left hand side (lhs)  
integral in Eq. (\ref{five}) from $\Delta x^4$ to $ V^4_U$.  
Collecting all the above results, from Eq. (\ref{five}) one obtains  
\hspace*{-0.5cm}\begin{eqnarray}  
%
\int\limits_{ V^4_U} \hspace*{-0.1cm}  d^4 x  
\frac{d^3 p}{p^0}~ \Theta_{in}~ p^\mu  ~{\partial}_\mu ~ \phi_{in} =   
\int\limits_{V^4_U}\hspace*{-0.1cm}  d^4 x  
\frac{d^3 p}{p^0}~ p^\mu n^*_\mu \times   
&&  
\nonumber \\  
\label{nine}  
%
[\phi_{in} - \phi_{out} ]~ \Theta(-p^\nu n^*_\nu)   
~\delta (t - t^*(\bar{x}) ) \,.&&  
\end{eqnarray}  
\vm  
\vm  
   
\noindent  
Since the  volumes $\Delta x^4$ and $V^4_U$ are arbitrary, one obtains   
the kinetic equation for the distribution function of the ``in'' domain   

\vm  
\begin{eqnarray}  
&& \Theta_{in}~ p^\mu  ~{\partial}_\mu ~ \phi_{in} (x,p) = C_{in} (x,p) +  \nonumber \\  
\label{ten}  
&&  p^\mu n^*_\mu [\phi_{in}(x,p) - \phi_{out}(x,p) ] ~\Theta(-p^\nu n^*_\nu)  
~\delta (t - t^*(\bar{x}) ) \,.  
\end{eqnarray}  
\vm  
\vm  
   
\noindent  
Note that the  general solution   of Eq. (\ref{nine})  
contains  
an arbitrary function $C_{in} (x,p)$  (the first term in the rhs of (\ref{ten}))  which identically vanishes while being 
integrated over the invariant momentum measure $ d^3 p /p_0 $.  
Such a property is  typical for a collision integral \cite{GROOT}, 
and we shall discuss its derivation  in the subsequent section.   
 
Similar one can obtain the equation for the distribution   
function of the ``out'' domain  

\vm    
\begin{eqnarray}  
&& \Theta_{out}~ p^\mu  ~{\partial}_\mu ~ \phi_{out} (x,p) = C_{out} (x,p) + \nonumber \\  
\label{eleven}  
&&  p^\mu n^*_\mu [\phi_{in}(x,p) - \phi_{out}(x,p) ]~ \Theta(p^\nu n^*_\nu)  
~\delta (t - t^*(\bar{x}) ) \,,  
\end{eqnarray}  
\vm  
\vm  
  
\noindent    
where the normal vector $n^*_\nu$ is  given by  (\ref{six}).  
Note the asymmetry between the rhs of Eqs. (\ref{ten})  
and (\ref{eleven}): for the space-like parts of hypersurface $\Sigma^*$   
the  source term with $\Theta(-p^\nu n^*_\nu) $ vanishes identically because   $p^\nu n^*_\nu > 0$.  
This reflects the  causal properties of the equations above:    
propagation of particles faster than light is forbidden, and hence no particle  
can (re)enter the ``in'' domain.   
  
\section{Collision Term for Semi-Infinite Domain.}  
  
\vm  
 
Since in the general case $\phi_{in}(x,p) \neq  \phi_{out}(x,p)$ on  $\Sigma^*$,  
the $\delta$-like terms in  the rhs of   
Eqs. (\ref{ten}) and (\ref{eleven}) cannot vanish simultaneously on this hypersurface.  
Therefore, the functions $\Theta_{in}^*  \equiv \Theta_{in}|_{\Sigma^*} \neq 0$ and   
$ \Theta_{out}^*  \equiv \Theta_{out}|_{\Sigma^*} \neq 0$ do not vanish   
simultaneously on  $\Sigma^*$ as well.   
Since there is no preference between ``in'' and ``out'' domains  
it is assumed that   
 
\vm  
\begin{equation}\label{twelve}  
\Theta_{in}^* = \Theta_{out}^* = \Theta (0) = \frac{1}{2}\,,  
\end{equation}  

\vm  
  
\noindent  
but the final results are independent of this choice.  
  
Now the collision terms for Eqs. (\ref{ten}) and (\ref{eleven}) can be readily obtained.    
Adopting the usual assumptions for   
the distribution functions \cite{BOGOL,GROOT,BALESCU}, one can   
repeat the standard derivation of the collision terms \cite{GROOT,BALESCU}  
and get the desired expressions.   
We shall not recapitulate this standard part, but only discuss  how to modify the derivation for   
our purpose.  
First of all, one has to start the derivation in the $\Delta x^4$ volume of  
the ``in'' domain and then  
extend it to the unified $4$-volume  
$V^4_U = \Delta x^4 \cup V^4_\Sigma$ similarly to the preceding section.  
Then the first part of the collision term for Eq. (\ref{ten}) reads  
\begin{eqnarray}\label{thirteen}  
C_{in}^{I} (x,p) & = & \Theta_{in}^2 \left( I_G [\phi_{in}, \phi_{in}] - I_L [\phi_{in}, \phi_{in}] \right)   
\,, \\  
\label{fourteen}  
I_G [\phi_{A}, \phi_{B}] & \equiv & \frac{1}{2} \int D^9 P~    
\phi_{A}(p^{\prime} )~ \phi_{B}(p_1^{\prime})~ W_{p\,p_1^{} | p^{\prime}p_1^{\prime}}   
\,, \\   
\label{fifteen}   
I_L [\phi_{A}, \phi_{B}] & \equiv & \frac{1}{2} \int D^9 P~  
\phi_{A}(p)~ \phi_{B}(p_1)~ W_{p\,p_1^{} | p^{\prime}p^{\prime}_1}\,,  
\end{eqnarray}  
  
\vm  
  
\noindent  
where the invariant measure of integration is denoted by   
$ D^9 P \equiv \frac{d^3 p_1}{p^0_1} \frac{d^3 p^{\prime} }{p^{\prime 0}}   
\frac{d^3 p^{\prime}_1 }{p^{\prime 0}_1} $ and $W_{p\,p_1^{} | p^{\prime}p^{\prime}_1}$   
is the transition rate in the elementary  reaction     
with energy-momentum conservation given in the form  
$p^\mu + p_1^\mu = p^{\prime \mu} + p^{\prime \mu}_1$.  
The rhs of  (\ref{thirteen}) contains the square of the  $\Theta_{in}$-function   
because the additional $\Theta_{in}$ accounts for the fact that   
on the boundary hypersurface $\Sigma^*$ one has to take  only one half   
of the traditional collision term (due to Eq. (\ref{twelve}) only one half of   
$\Sigma^*$ belongs to the ``in'' domain).   
It is easy to understand that  
on $\Sigma^*$   
the second part of the collision term   
(according to Eq. (\ref{twelve})) is defined by the   
collisions between particles of ``in'' and ``out'' domains  
\begin{equation}\label{sixteen}  
\hspace*{-0.25cm}  
C_{in}^{II} (x,p)  =  \Theta_{in} \Theta_{out}  \left( I_G [\phi_{in}, \phi_{out}] - I_L [\phi_{in}, \phi_{out}] \right)  
.   
\end{equation}  

\noindent  
Combining (\ref{ten}), (\ref{thirteen}) and (\ref{sixteen}), one gets the kinetic  
equation for the  
``in'' domain  
\begin{eqnarray} 
&& \Theta_{in}~ p^\mu  ~{\partial}_\mu  \phi_{in} (x,p) =  C_{in}^{I} (x,p) +  
C_{in}^{II} (x,p) + p^\mu n^*_\mu \times \nonumber \\ 
\label{seventeen} 
%
&&  [\phi_{in}(x,p) - \phi_{out}(x,p) ] ~\Theta(-p^\nu n^*_\nu) 
~\delta (t - t^*(\bar{x}) ) \,. 
\end{eqnarray} 
\vm 
\vm 
\vm 
 
\noindent  
 
The kinetic equation for the ``out'' domain  
can be derived similarly and then it can be   
represented in the form  
\begin{eqnarray} 
&& \Theta_{out}~ p^\mu  ~{\partial}_\mu  \phi_{out} (x,p) =  C_{out}^{I} (x,p) +  
C_{out}^{II} (x,p) + p^\mu n^*_\mu \times \nonumber \\ 
\label{eighteen} 
&&  [\phi_{in}(x,p) - \phi_{out}(x,p) ] ~\Theta(p^\nu n^*_\nu) 
~\delta (t - t^*(\bar{x}) ) \,, 
\end{eqnarray} 
where the evident notations for the collision terms 
$C_{out}^{I} \equiv \Theta_{out}^2 \left( I_G [\phi_{out}, \phi_{out}] - I_L [\phi_{out}, \phi_{out}] \right) $  
and  
$C_{out}^{II} \equiv \Theta_{in} \Theta_{out} \left( I_G [\phi_{out}, \phi_{in}] - I_L [\phi_{out}, \phi_{in}] \right) $ 
are used.  
  
The equations  (\ref{seventeen}) and (\ref{eighteen}) can be represented also 
in a covariant form with the help of the function ${\cal F}^{*}(t, \bar{x})$.  
Indeed, 
applying the definition of the derivative of the implicit function to  
$\partial_\mu t^*(\bar{x}) $, 
one can rewrite the external normal vector (\ref{six}) as  
$n^*_\mu \equiv \partial_\mu {\cal F}^{*}(t, \bar{x}) / \partial_0 {\cal F}^{*}(t, \bar{x})$. 
Now using the inequality $\partial_0 {\cal F}^{*}(t^{*}, \bar{x}) > 0$  and  
the following identities  
$\delta ( {\cal F}^{*}(t, \bar{x}) ) = \delta (t - t^*(\bar{x}) ) /   
\partial_0 {\cal F}^{*}(t^*, \bar{x})$,  
$\Theta_{A} \equiv \Theta ( S_A~{\cal F}^{*} (t, \bar{x})  )$  
one can write Eqs. (\ref{seventeen}) and (\ref{eighteen})   
in a fully  covariant way  
\begin{eqnarray} 
&& \Theta_{A}~ p^\mu  ~{\partial}_\mu ~ \phi_{A} (x,p) =  C_{A}^{I} (x,p) +  
C_{A}^{II} (x,p) + p^\mu \partial_\mu {\cal F}^{*} \times 
\nonumber \\ 
\label{nineteen} 
&&  [\phi_{in}(x,p) - \phi_{out}(x,p) ] ~\Theta(S_A~p^\nu \partial_\nu {\cal F}^{*}) 
~\delta ( {\cal F}^{*}(t, \bar{x})  ) \,, 
\end{eqnarray} 
where the notations $A \in in$, $S_{in}=-1$ ($A \in out$,  $S_{out}=1$)  
are introduced for ``in'' (``out'') domain.

For the continuous distribution functions on $\Sigma^*$,  
i.e. $\phi_{out}|_{\Sigma^*} = \phi_{in}|_{\Sigma^*}$,     
the source terms on rhs of Eqs. (\ref{seventeen})    
and (\ref{eighteen})  
vanish and one  
recovers the Boltzmann equations.  
Moreover, with the help of the evident relations  
%
\begin{eqnarray}\label{twenty}  
&&- {\partial}_\mu ~ \Theta_{in} = {\partial}_\mu ~ \Theta_{out} =  
~\delta ( {\cal F}^{*}(t, \bar{x})  ) ~\partial_\mu {\cal F}^{*} (t, \bar{x}) \,, \\ 
\label{twone}  
&&C_{in}^{I} + C_{in}^{II} + C_{out}^{I} + C_{out}^{II} =    
I_G [\Phi, \Phi] - I_L [\Phi, \Phi]\,,   
\end{eqnarray}   

\noindent  
where  
$\Phi(x,p) \equiv \Theta_{in}~\phi_{in}(x,p) + \Theta_{out}~\phi_{out}(x,p) $,   
one can get the following result summing up  Eq. (\ref{seventeen}) and (\ref{eighteen})   
\begin{equation}\label{twtwo}  
p^\mu  ~{\partial}_\mu ~ \Phi (x,p) = I_G [\Phi, \Phi] - I_L [\Phi, \Phi]\,.  
\end{equation}  

\noindent  
In other words, the usual Boltzmann equation follows from  
the system (\ref{nineteen})    
automatically  {\it  without any assumption} about the behavior   
of $\phi_{in}$ and $\phi_{out}$ on the boundary hypersurface  
$\Sigma^*$.  
Also  Eq. (\ref{twtwo}) is valid    
not only under condition (\ref{twelve}),  
but for {\it any choice}  $0 < \Theta_{A}^{*} < 1$ obeying Eq. (\ref{eight}).  

In fact the system (\ref{nineteen}) generalizes   
the relativistic kinetic equation to the case of the strong   
temporal and spatial inhomogeneity, i.e.,  
for  $\phi_{in}(x,p) \neq \phi_{out}(x,p)$ on $\Sigma^*$.  
Of course, one has to be extremely careful while discussing  
the strong temporal inhomogeneity (or discontinuity on the space-like parts of $\Sigma^*$)   
such as the so called {\it time-like shocks} \cite{TIMESHOCKa,TIMESHOCKb}  
because, as shown in the subsequent section, their existence  
contradicts  the usual assumptions \cite{BOGOL,GROOT,BALESCU}    
adopted for distribution functions.   
  
From the system (\ref{nineteen}) it is possible to derive the   
macroscopic equations of  motion for the  energy-momentum tensor
 by multiplying the corresponding equation with $p^\nu$   
and integrating it over the invariant measure. Thus,  Eq. (\ref{nineteen})   
generates the following expression  
($T^{\mu \nu}_{A} \equiv \int \frac{d^3 p }{p^ 0}~ p^\mu p^\nu \phi_{A}(x,p)$)  

\vm  
\begin{eqnarray}  
&&\Theta_{A}~ {\partial}_\mu ~ T^{\mu \nu}_{A}  =   \int \frac{d^3 p }{p^ 0}~  p^\nu    
C_{A}^{II} (x,p) + \int \frac{d^3 p }{p^ 0}~ p^\nu   
p^\mu \partial_\mu {\cal F}^{*} \times \nonumber \\  
\label{twthree}  
&&[\phi_{in} (x,p) - \phi_{out} (x,p) ] ~\Theta(S_A~p^\rho \partial_\rho {\cal F}^{*})  
~\delta ( {\cal F}^{*}(t, \bar{x})  ) .  
\end{eqnarray}  
\vm  
\vm  
   
  
\noindent  
Similar to the usual Boltzmann equation  
the momentum integral of the collision term $C_{in}^{I}$ vanishes  
due to its symmetries \cite{GROOT},  
but it can be shown   
that the integral of the second collision term $C_{in}^{II}$ does     
not vanish because it involves two different 
distribution functions.   
 
The corresponding system of  equations  for the conserved current   
$N^\mu_A \equiv \int \frac{d^3 p }{p^ 0}~ p^\mu \phi_{A} (x,p)  $  
can be obtained by  direct integration of the system (\ref{nineteen})    
with the invariant measure  
\vm 
\begin{eqnarray} 
\Theta_{A}~ {\partial}_\mu ~ N^{\mu}_{A} & = &      
\int \frac{d^3 p }{p^ 0}~   p^\mu \partial_\mu   {\cal F}^{*}  
[\phi_{in} (t, \bar{x}) - \phi_{out} (t, \bar{x}) ] \times  \nonumber \\ 
\label{twfour} 
&&\Theta(S_A~p^\rho \partial_\rho  {\cal F}^{*} ) 
~\delta ( {\cal F}^{*}(t, \bar{x})  ) . 
\end{eqnarray} 
\vm 
  
\noindent 
The  above equation does not contain the contribution from antiparticles 
(just for simplicity), but  the latter  can be easily recovered.     
Note that  in contrast to (\ref{twthree}) the momentum integral 
of both collision terms vanish in  Eq. (\ref{twfour}) due to symmetries.   
    
   
 
\section{Conservation Laws at $\Sigma^*$ }  
  
It is clear that  Eqs. (\ref{nineteen}), (\ref{twthree}) and (\ref{twfour})   
remain   
valid  both for finite domains and   
for a  multiple valued function $t = t^*(\bar{x})$ as well.  
To derive the whole system of these equations  
in the  latter case,  one has to divide the function  
$t^*(\bar{x})$ into the single valued parts, but this discussion is beyond the scope of this paper.  
Using Eqs.  
(\ref{nineteen}), (\ref{twthree}) and (\ref{twfour})  
we are ready to analyze the boundary conditions  
on the hypersurface $\Sigma^*$.  
The simplest way to get the boundary conditions 
is to integrate  Eqs. (\ref{twthree}) and (\ref{twfour}).  
Indeed, integrating (\ref{twthree})  
over the 4-volume $V^4_{p \Sigma}$ (shown as the area ABCD in Fig. 2) containing part 
$p \Sigma^*$ of the 
hypersurface  $\Sigma^*$, 
one obtains the energy-momentum conservation.   
Before 
applying the Gauss theorem to the lhs of (\ref{twthree}),  
we  note that the corresponding $\Theta_A$-function reduces 
the 4-volume $V^4_{p \Sigma}$ to its part which belongs to  the $ A$-domain. 
The latter is  
shown as area $ALMD$ ($BCML$) for $A \in in$ ($A \in out$) in Fig. 2.  
Then 
in the limit of a vanishing maximal distance  $\Delta \rightarrow 0$ 
between the hypersurfaces $AD$ and $BC$ in Fig. 2, 
the volume integral of the lhs of Eq. (\ref{twthree}) 
can be rewritten as the two integrals  $\int d\sigma_\mu T^{\mu\nu}_A$ : the first  integral is performed over the hypersurface $p\Sigma$ 
shown  as an arc 
$LM$ in Fig. 2,  and the second  integral  reduces to the same hypersurface  but taken in the opposite direction, i.e. the $ML$ arc in Fig. 2. 
Thus, the volume integral of the lhs of Eq. (\ref{twthree}) vanishes in this limit, and 
we obtain 
\begin{eqnarray} 
&& 0 \hspace*{-0.05cm} = \hspace*{-0.1cm} \int\limits_{ V^4_{p \Sigma} } \hspace*{-0.1cm}  d^4 x~\Theta_A 
{\partial}_\mu  \lp  T^{\mu \nu}_A (x,p) \rp \hspace*{-0.01cm}  \equiv \hspace*{-0.1cm} 
\int\limits_{ V^4_{p \Sigma} } \hspace*{-0.1cm}
  d^4 x~ \frac{d^3 p }{p^ 0}~  p^\nu   C_{A}^{II} (x,p) +   
\nonumber \\  
&&  
\int\limits_{ V^4_{p \Sigma} }  d^4 x~  \frac{d^3 p }{p^ 0}~ \delta ( {\cal F}^{*}(t, \bar{x})  )~  
p^\nu p^\mu \partial_\mu {\cal F}^{*}~ \times  \nonumber \\ 
\label{twfive} 
&&  
[\phi_{in} (x,p)   - \phi_{out}(x,p) ] ~ \Theta(S_A~p^\rho \partial_\rho  {\cal F}^{*})   \,.  
\end{eqnarray} 
\vm 
 
\noindent 
Similarly  to Sect. II, in the limit $\Delta \rightarrow 0$ the second integral on the  rhs  
of (\ref{twfive}) can be reexpressed as    
an integral over the closed hypersurface.  
Since the latter  is arbitrary, then Eq. (\ref{twfive}) 
can be satisfied, if and only if the  energy-momentum conservation occurs  
for every point of the hypersurface $\Sigma^*$

\begin{eqnarray}
\label{twsix}
&&
T^{\mu \nu}_{in \pm} ~
\partial_\mu {\cal F}^{*}(t^{*}, \bar{x})  =
T^{\mu \nu}_{out \pm} ~
\partial_\mu {\cal F}^{*}(t^{*}, \bar{x})  \,, \\
&&
 T^{\mu \nu}_{A \pm}  \equiv
\int \frac{d^3 p }{p^ 0}~ p^\mu p^\nu \phi_{A} (x,p)~
\Theta(\pm~p^\rho \partial_\rho {\cal F}^{*})
 \,.  \nonumber
\end{eqnarray}
\vm

\noindent 
In deriving (\ref{twsix}) from (\ref{twfive})  
we used the fact that  
the 4-volume  integral of the second collision term  $C_{A}^{II}$   
vanishes for finite values of distribution functions  
because of the Kronecker symbols.  
The results for the conserved current follows similarly from Eq. (\ref{twfour}) 
after integrating it  
over the 4-volume $V^4_{p \Sigma}$ and taking the limit $\Delta \rightarrow 0$

\begin{eqnarray}
\label{twseven}
&&
 N^{\mu }_{in \pm}  ~
\partial_\mu {\cal F}^{*}(t^{*}, \bar{x})  =
 N^{\mu }_{out \pm}  ~
\partial_\mu {\cal F}^{*}(t^{*}, \bar{x}) \,, \\
%
%
%
&&
 N^{\mu }_{A \pm}  \equiv
\int \frac{d^3 p }{p^ 0}~ p^\mu \phi_{A} (x,p)~ \Theta(\pm~p^\rho \partial_\rho {\cal F}^{*})
 \,.  \nonumber
\end{eqnarray}
\vm

\noindent 
The fundamental  difference  between the conservation laws (\ref{twsix}),  (\ref{twseven})  
and the ones of  usual hydrodynamics  is that the systems (\ref{twsix}) and   (\ref{twseven}) 
conserve the quantities of the outgoing from ($S_A = 1$)   and  incoming  
to ($S_A = -1$) ``in'' domain particles 
{\it separately}, whereas in usual hydrodynamics only the  sum of these  contributions 
is conserved.    
 
\vspace*{0.3cm}

\begin{figure}
\hspace*{0.0cm}\mbox{ \epsfig{file=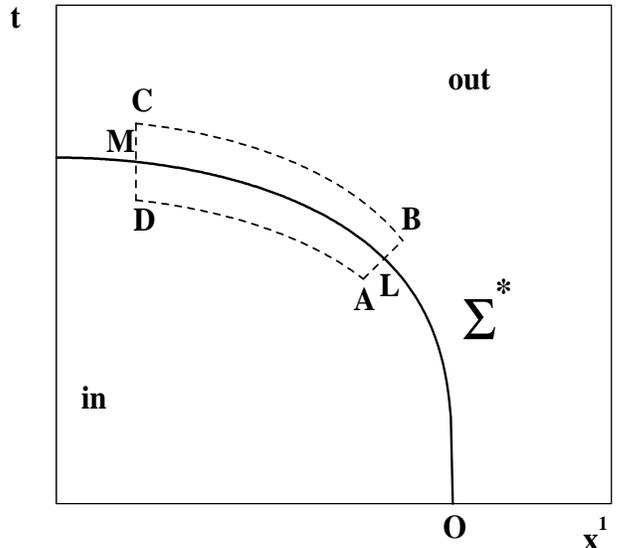,width=80mm,height=75mm} }
\caption{
Schematic two dimensional picture
of the integration contour to derive the boundary conditions (25) - (27) between the ``in'' and
``out'' domains. In the limit of  a vanishing maximal distance
 $\Delta  \rightarrow 0$ between the hypersurfaces $AD$ and $BC$,  both of these
hypersurfaces are reduced to the part $p\Sigma^*$ (an arc $LM$) of the boundary
 $\Sigma^*$ between domains.
}
\label{fig2}
\end{figure}

\vspace*{-0.2cm}

The trivial solution of Eqs. (\ref{twsix}) and (\ref{twseven}) corresponds to a continuous  
transition between ``in'' and ``out'' domains 
 
\begin{equation}\label{tweight}  
\phi_{out} (x, p) \Bigl|_{\Sigma^*} = \phi_{in} (x, p) \Bigr|_{\Sigma^*}\,.   
\end{equation}

This choice  corresponds to the BD model \cite{BD:00}.   
The  BD model gives  a correct result for an oversimplified kinetics 
considered here. However, in the  case of the first order  phase  transition  
(or a strong cross-over) which was a prime  target of the hydro+cascade  models     
 \cite{BD:00,SHUR:01} the situation is different.  
In the latter case the speed of sound either vanishes (or becomes very small) \cite{Hung:94,Hung:98} 
and, hence, the rarefaction shock waves become  possible \cite{RAREF:1,RAREF:2,BUG:88}.    
The reason why the rarefaction shocks  may exist lies in the anomalous thermodynamic 
properties \cite{BUG:88} of the media near the phase transition region.    
In other words,  on the boundary between the  mixed  and  
hadronic phases the rarefaction shocks are mechanically 
stable  \cite{BUG:88}, whereas the compression shocks are mechanically unstable. 
This is also valid for the vicinity of  the generalized mixed phase of 
a strong cross-over.
 
One important  consequence of the shock  mechanical stability criterion  
is that the stable  shocks necessarily are supersonic in the media where they propagate.  
The latter means that the continuous rarefaction flow  
in the region of phase transition  is mechanically unstable as well,  
since  a rarefaction shock, if  it appears, 
propagates inside the fluid faster than the sound wave and, hence,  
it should change the fluid's state.    
Due to this reason   the unstable hydrodynamic solutions simply do not appear \cite{HYDRO,BUG:89}.

Applying these arguments to the BD model, one concludes:   
for the first order phase transition or strong cross-over the sound wave in 
the (generalized) mixed phase may be unstable and the strong discontinuities  
of the thermodynamic quantities  
are possible \cite{RAREF:1,RAREF:2,BUG:88}. The latter corresponds to   
the   non-trivial solution of the conservation 
laws (\ref{twsix}) and (\ref{twseven}), which allows    
a discontinuity of the   
distribution function on two sides of the hypersurface $\Sigma^*$.  
Since there is twice the number of  conservation laws  compared to the  usual 
hydrodynamics,   
it is impossible, as shown below,   to build up  
the nontrivial solution of Eqs. (\ref{twsix}) and (\ref{twseven}),  
if the distribution functions on both sides of the hypersurface $\Sigma^*$,  i.e. $\phi_{in}$ and $\phi_{out}$, are    taken to be 
the equilibrium ones.

Consider first the space-like parts of the hypersurface $\Sigma^*$. 
Then Eqs. (\ref{twsix}) and (\ref{twseven}) for $S_A = -1$ vanish identically 
because of  the inequality $p^\mu \partial_\mu {\cal F}^{*}(t^{*}, \bar{x}) > 0$, 
whereas for $S_A = 1$ Eqs. (\ref{twsix}) and (\ref{twseven})  recover   
the usual hydrodynamical conservation laws at  the discontinuity.   
However,  
it can be shown that the existence of  strong discontinuities across the space-like hypersurfaces,  
the {\it time-like shocks} \cite{TIMESHOCKa,TIMESHOCKb},  
is rather  problematic because it leads to a contradiction  of  the basic 
assumptions adopted for the distribution function,   
even though  the conservation 
laws (\ref{twsix}) and (\ref{twseven}) are formally fulfilled.

Indeed, according to the Bogolyubov's classification \cite{BOGOL},  
a one-particle treatment can be established for   
a   typical time $\Delta t$ 
which, on one hand,   should be much larger than  
the collision time $\tau_{Coll}$,  and, on the other hand,  
it should be much smaller than the relaxation time $\tau_{Relax}$  
 
\begin{equation}\label{twnine} 
\tau_{Coll} << \Delta t << \tau_{Relax}\,.  
\end{equation} 

\noindent 
Similar to the usual Boltzmann equation (see also discussions in \cite{BOGOL,BALESCU}), 
in  deriving the collision  terms of  Eq. (\ref{nineteen}) we implicitly   
adopted the requirement that 
the distribution function does not change  substantially for times $\Delta t$ 
less than the relaxation time $\tau_{Relax}$. 
However, at  the discontinuities on the space-like parts of $\Sigma^*$, 
suggested in  \cite{TIMESHOCKa,TIMESHOCKb},   
the distribution function changes suddenly, i.e. $\Delta t = 0$,   
and  
the left inequality (\ref{twnine}) cannot be fulfilled 
at the {\it time-like shock}. 
Therefore, according to the Bogolyubov's classification \cite{BOGOL},  such  a process,   
 which is 
shorter than the typical collision time,  belongs to a  prekinetic or chaotic stage
and, hence,  cannot be studied  at  the level of a one-particle distribution 
function. 
It would instead  require  the analysis of  a  hierarchy of  $N$-particle   
distribution functions, where $N$ is the number of particles in the system.     
Thus, the existence of time-like shocks contradicts  the 
adopted assumptions  for a one-particle distribution.
Their existence  should be  demonstrated  first 
within the higher order  distributions.  
This  statement  applies  to several  papers published by the
Bergen group during the last few years where   {\it time-like shocks }  were
attenuated  in time  using a  phenomenological quasi-kinetic approach \cite{Laszlo:04}.
For  the same reason, the use of equilibrium values  for temperature and chemical potential 
in an attenuated time-shock is rather problematic for time scales shorter than $ \tau_{Coll}$.
Note, however, that the discontinuities at  the time-like parts of $\Sigma^*$
(usual shocks)  have no such restrictions and, hence, in what follows we shall analyze  
only   these discontinuities.

 
\section{Boundary Conditions at $\Sigma^*$ for a Single Degree of Freedom}

Now  we have   to  find out  whether it is  principally possible 
to obtain 
the nontrivial solution of systems  (\ref{twsix}) and (\ref{twseven}) 
using the parts of  equilibrium distributions on  
the time-like segments  of the hypersurface $\Sigma^*$.  
%
To simplify  the presentation,  first 
we consider  the same kind of particles in both domains.
As usual in relativistic hydrodynamics   it is convenient to transform  the coordinate system  
$(t^*(\bar x ); \bar x )$ into the special local frame
$(t^*_L (\bar x_L ); \bar x_L )$ being, for definiteness, 
the rest-frame of  discontinuity   between the distributions $\phi_{in}$ and $\phi_{out}$  and  defined as follows:   
the $x$-axis  should coincide with the local external normal  vector 
to the hypersurface $\Sigma^*$, whereas $y$- and $z$-axes should  
belong to  the tangent hyperplane to $\Sigma^*$.   
In this case the external  normal vector to the time-like parts of $\Sigma^*$ has a simplest form: 
$n^*_\mu = (~0 ;~ \partial_1 {\cal F}^{*}_L; ~0; ~0)$, 
and one can readily check 
that the value of the derivative  $ \partial_1 {\cal F}^{*}_L$ plays
an important   role 
in the conservation laws (\ref{twsix}) and (\ref{twseven}) 
only through the  the {\it cut-off} $\Theta$ function.
Then, like in the theory of  usual relativistic shocks \cite{HYDRO,BUG:88,BUG:89}, 
it can be shown that equations for  $y$- and $z$-components of system (\ref{twsix})     
degenerate to the identities because of the symmetries  of the  energy-momentum tensor. 
Therefore, the number of   independent  equations at the discontinuity is  7:  
a switch off criterion  
and six independent equations out of systems (\ref{twsix}) and (\ref{twseven}) 
($t$- and $x$-equations (\ref{twsix}) and one equation (\ref{twseven}) for 
two choices of $S_A = \{ -1; +1\}$).       
 
On the other hand the number of unknowns  is 6 only: temperature $T^*_{in}$ and  
baryonic chemical potential $\mu^*_{in}$ of the ``in'' domain,  
temperature $T^*_{out}$ and 
baryonic chemical potential $\mu^*_{out}$ of the ``out'' domain,   the collective velocity $v^*_{in}$
of the ``in'' domain  particles,
and the collective velocity $v^*_{out}$ of the particles of ``out'' domain,  
which, evidently,  should be collinear   
to the normal vector $n^*_\mu$ in the rest-frame of the discontinuity.   
Thus, a formal counting of equations and unknown shows that it is impossible 
to satisfy the conservation laws (\ref{twsix}) and (\ref{twseven}), if  
the distribution functions on both sides are the  equilibrium ones.   
 
The last  result means that instead of  a traditional 
discontinuity we have to search for a principally new boundary conditions  
on the  hypersurface $\Sigma^*$. The analysis  shows that there are two of such possibilities 
with the equilibrium  distribution function in the ``in'' domain and  
a special 
superposition 
of  
two  {\it cut-off} equilibrium  distributions for the ``out'' domain.    
The first possibility is to choose $\phi_{out}$  as follows: 
\begin{eqnarray}\label{thirty} 
\hspace*{-0.9cm}\phi_{out} \Bigl|_{\Sigma^*}  \hspace*{-0.2cm} & = &   
\phi_{in}~ (~T_{in}^*, ~\mu_{in}^*, ~v^*_{in} )~~ \Theta(~~ p^1  \partial_1 {\cal F}^*_L )  +    
\nonumber   \\ 
\hspace*{-0.9cm}
& & \phi_{out} (T_{out}^*, \mu_{out}^*, v_{out}^*)~ \Theta(- p^1  \partial_1 {\cal F}^*_L )   
 \, ,   
\end{eqnarray} 
i.e. the distribution of outgoing particles from the ``in'' domain (the first term in the rhs of
Eq. (\ref{thirty})) is continuous on the hypersurface $\Sigma^*$,
whereas  the distribution  of the incoming to ``in'' domain  particles (the second term in  
the rhs of Eq. (\ref{thirty})) 
has a discontinuity on $\Sigma^*$ which, however,  
conserves the energy, momentum and baryonic charge because of
the following boundary conditions
 ($ \nu = \{0; ~1\}$) 
\begin{eqnarray} 
%
&&
T^{1 \nu}_{in -}~\, (T_{in}^*, \mu_{in}^*, v_{in}^*) ~ =~
\label{thone}
T^{1 \nu}_{out -}~(T_{out}^*, \mu_{out}^*, v_{out}^*)  \,, \\
%
%
&&
N^{1 }_{in -} ~ (T_{in}^*, \mu_{in}^*, v_{in}^*) ~ =~
N^{1 }_{out -}~(T_{out}^*, \mu_{out}^*, v_{out}^*)
\label{thtwo} 
\, . 
\end{eqnarray} 
\vm 
  
The above choice  of boundary conditions at  $\Sigma^*$  
allows to reduce the systems (\ref{twsix}) and (\ref{twseven})  
for $S_A = 1$ to the identities, and, hence,   
from the systems (\ref{twsix}) and (\ref{twseven}) 
there remain only   three independent  equations   
(\ref{thone}), (\ref{thtwo}) for $S_A = - 1$.   
Alone with a  switch off criterion,  these four equations can be     
solved now  for six  independent variables, and, consequently, 
the two variables can be chosen free for a moment.  
Thus, we showed that  both  the outgoing and incoming parts of  
the distribution function (\ref{thirty}) 
can be chosen  as the equilibrium ones, but with different   
temperatures, chemical potentials and  non-zero relative velocity
 $v^*_{rel}  \equiv ( v^*_{out} - v^*_{in} ) / ( 1 -  v^*_{out}  v^*_{in} ) $
with respect to the distribution function $\phi_{in}$.  

Note a principal difference between this discontinuity  and all ones known  before:
the  ``out'' domain state consists, in general,  of two  different subsystems (fluxes) that have
their individual hydrodynamic parameters. It is clear
that  it is impossible to reduce  three of those hydrodynamical parameters of one flux to 
those three of other flux 
because there are only two free variables out of six. 
Thus, together with the ``in'' domain flux  there are
in total  three fluxes involved  in this discontinuity, and,  therefore, it is appropriate to
name it a {\it three flux discontinuity} in order to distinguish it from the ordinary shocks
that are defined  maximum by two fluxes.

The outgoing component of the distribution (\ref{thirty}) 
coincides with the  choice of the boundary conditions 
suggested in  the  TLS model \cite{SHUR:01}, whereas the 
equations (\ref{thone}) and (\ref{thtwo}) are missing in this model. 
This is the  reason why the TLS model suffers from the energy, momentum and 
charge non-conservation.  It is also necessary to note that 
the lower values  of the  temperature $T_{out}^* \le T_{in}^* $ and 
baryonic  chemical potential $\mu_{out}^* \le \mu_{in}^* $,   
which are typical for the rarefaction process considered in \cite{SHUR:01},  
should be compensated by an  extra flow from  the incoming particles to  
the ``in'' domain, i.e. $v_{rel}^*$ should be opposite to the external normal vector 
$n_\mu^*$ in the rest-frame of the  {\it three flux  discontinuity}.
Therefore, such a discontinuity is analogous 
to the compression shock wave in relativistic hydrodynamics,  
 and, hence, it  
cannot appear in  the rarefaction process for    
any of the hadronic species considered in Ref. \cite{SHUR:01}.

Similarly, one can find  
another non-trivial  solution of the systems (\ref{twsix}) and (\ref{twseven}) 
which   
corresponds to  opposite choice to Eq. (\ref{thirty})
\begin{eqnarray}\label{ththree} 
\hspace*{-0.9cm} 
\phi_{out} \Bigl|_{\Sigma^*} \hspace*{-0.2cm}  & = &   
\phi_{in}~ (~T_{in}^*, ~\mu_{in}^*, ~v^*_{in} )~~ \Theta(- p^1  \partial_1 {\cal F}^*_L )  +    
\nonumber   \\ 
\hspace*{-0.9cm}
& & \phi_{out} (T_{out}^*, \mu_{out}^*, v_{out}^*)~ \Theta(~~ p^1  \partial_1 {\cal F}^*_L )   
 \, ,   
\end{eqnarray} 
i.e., the incoming to the ``in'' domain component of the 
distribution above (the first  term  in the rhs of Eq. (\ref{ththree})) 
is continuous on hypersurface $\Sigma^*$, but 
the outgoing from the ``in'' domain component has a  
discontinuity on $\Sigma^*$ which  obeys  
the following conservation laws ($ \nu = \{0; ~1\}$):   
\begin{eqnarray} 
%
&&
T^{1 \nu}_{in +}~\, (T_{in}^*, \mu_{in}^*, v_{in}^*) ~ =~
\label{thfour}
T^{1 \nu}_{out +}~(T_{out}^*, \mu_{out}^*, v_{out}^*)  \,, \\
%
%
&& 
N^{1 }_{in +} ~ (T_{in}^*, \mu_{in}^*, v_{in}^*) ~ =~
N^{1 }_{out +}~(T_{out}^*, \mu_{out}^*, v_{out}^*)
\label{thfive} 
\, . 
\end{eqnarray} 
It is  clear that both the outgoing and incoming components of the 
distribution (\ref{ththree}) can be chosen as the equilibrium distribution  
functions. A simple analysis of the system (\ref{thfour}), (\ref{thfive}) 
shows that for $T_{out}^* \le T_{in}^*$ and $\mu_{out}^* \le \mu_{in}^*$
the relative velocity $v_{rel}$  in the local  
frame should be collinear to the external normal vector $n_\mu^*$, and, hence, 
such a discontinuity is analogous to the rarefaction 
shock wave in the relativistic hydrodynamics.  
 Thus,  in contrast to  the  TLS choice, Eq. (\ref{ththree}) should 
be used  as the initial conditions for  the ``out'' domain 
while studying  the rarefaction process of  
 matter  with anomalous thermodynamic properties. 
 
Now we are ready to discuss the question  how  
the non-trivial solutions (\ref{thirty}) and (\ref{ththree})  
will modify the system of the hydro+cascade equations  
(\ref{nineteen}), (\ref{twthree}) and (\ref{twfour}).   
In what follows we shall assign  the hydrodynamic equations 
to the  ``in'' domain and the cascade  ones to the 
``out'' domain (the opposite case 
can be considered similarly).   
Applying Eqs. (\ref{thirty}), (\ref{thone}) and (\ref{thtwo})   
to the ``in'' Eqs. (\ref{twthree}) and (\ref{twfour})  
and to the ``out'' Eq. (\ref{nineteen}), one obtains the following  
system: 
\begin{eqnarray} 
\label{thsix} 
&&\Theta_{in}~ {\partial}_\mu ~ T^{\mu \nu}_{in}  =   \int \frac{d^3 p }{p^ 0}~  p^\nu 
C_{in}^{II} (x,p) \, ,  \\   
\label{thseven} 
&& \Theta_{in}~ {\partial}_\mu ~ N^{\mu}_{in} = 0 \, , \\ 
\label{theight} 
&& \Theta_{out}~ p^\mu  ~{\partial}_\mu ~ \phi_{out} (x,p) =  C_{out}^{I} (x,p) + 
C_{out}^{II} (x,p) \, ,   
\end{eqnarray} 
i.e., due to the boundary conditions (\ref{thirty}) -- (\ref{thtwo})   
the $\delta$-like terms have disappeared from the original system of equations.  
Also it is clear that the source  term in the rhs of Eq. (\ref{thsix}) 
does not play any role because it is finite on the hypersurface $\Sigma^*$ 
and it  vanishes everywhere outside of $\Sigma^*$.   
 
In order to obtain the system of hydro+cascade equations  
(\ref{thsix}) -- (\ref{theight}) 
for the non-trivial solution defined by Eqs. (\ref{ththree}) -- (\ref{thfive}), 
the hydrodynamic description  
has to be extended  
to the  outer $\varepsilon$-vicinity  
($\varepsilon \rightarrow 0$)  of the 
hypersurface $\Sigma^*$  
\begin{eqnarray} 
\label{thnine} 
&&\Theta_{out}~ {\partial}_\mu ~ T^{\mu \nu}_{out}  =   \int \frac{d^3 p }{p^ 0}~  p^\nu 
C_{out}^{II} (x,p) \, ,  \\ 
\label{forty} 
&& \Theta_{out}~ {\partial}_\mu ~ N^{\mu}_{out} = 0 \, ,  
\end{eqnarray} 
which  in practice  means that for Eqs. (\ref{ththree}) -- (\ref{thfive}) 
one has to start solving the cascade equation 
(\ref{theight}) just bit inside of the ``out'' domain   
in order to get rid of the  $\delta$-like term in  (\ref{theight}) and 
move this term  to the discontinuity on the hypersurface $\Sigma^*$.  
 
The remarkable feature of the system of hydro+cascade equations  
(\ref{thsix}) -- (\ref{forty}) is that   
each equation automatically vanishes outside of the domain where it 
is specified. Also, by the construction, it is free of the principal 
difficulties of the BD and TLS models discussed above.   
The question how to conjugate the {\it three flux discontinuity} with the solution of the hydro 
equations  (\ref{thsix}), (\ref{thseven}), (\ref{thnine}) and (\ref{forty}) will be discussed
in the next section.


 
\section{Boundary Conditions at $\Sigma^*$ for Many Degrees of Freedom}

In order to apply the above results to 
the description of the QGP-HG phase transition that occurs in relativistic 
nuclear collisions  
it is necessary to take into account the fact that the real situation differs 
from the previous consideration in two respects. 
The first one is that  in the realistic case inside  the ``in '' domain there should  
exist the QGP, whereas it should not appear in the ``out'' domain.   
Of course, the discussion of the QGP kinetic theory is much more complicated problem 
and it lies far beyond the subject of this work.  
For our purpose it is sufficient to generalize the equations of motion (\ref{thsix}) - (\ref{forty})  inside domains 
and the conservation laws (\ref{twsix}) and (\ref{twseven}) between these  domains to the realistic case. 
Such a generalization can be made  because in the case of the QGP-HG phase transition
there will be also an exchange
of particles between  the ``in'' and ``out''  domains which has to be accounted by the $\delta$-like
source terms in the transport equations. The only important difference from the formalism
developed in the preceding   sections is  that QGP must   hadronize
while entering 
the ``out'' domain, whereas the  hadrons should  melt while entering the ``in'' domain. 
Note, however, that in relativistic hydrodynamics one has 
to assume that all reaction, i.e.  the QGP hadronization and melting of hadrons  in this case,
occur instantaneously.  Under this assumption  
one can justify the validity of 
the equations of motion (\ref{thsix}) - (\ref{forty})  and the conservation laws between
QGP and HG on the boundary $\Sigma^*$.
 
The second important fact to be taken into account is that some hadrons  
have the large scattering cross-sections with other particles and some hadrons  
have the small cross-sections, and because of that   the hadrons of both kinds  participate in the  
collective flow differently.  
A  recent effort \cite{BUGGG:01,BUG:02B} to classify the inverse slopes of the  
hadrons at SPS lab energy  158 GeV$\cdot$A      
led to the  conclusion that   
the most abundant  hadrons like  pions, kaons,  (anti)nucleons, 
$\Lambda$ hyperons e.t.c. participate in the hadron  rescattering and resonance decay 
till the very late time of expansion,  
whereas $\Omega$ hyperons, $J/\psi$ and $\psi^\prime$ mesons    
practically do not interact with the  hadronic media and, hence,    
the freeze-out of their transverse  momentum spectra  ({\it kinetic freeze-out}) may occur just at 
hadronization temperature $T_H$. Therefore,  
the inverse slopes of the $\Omega$, $J/\psi$ and $\psi^\prime$ particles 
is a combination of the thermal motion and the transversal expansion of 
the media from which those particles are formed.   
 
These results for the $\Omega$ baryons and $\phi$ mesons were obtained  
within the BD and TLS models, whereas for the $J/\psi$ and $\psi^\prime$ mesons 
it was suggested in Refs.  \cite{BUGGG:01,BUG:02B} for the first time.   
Later on  these results  
were refined further in Ref. \cite{GBG:02}  by the simultaneous fit with 
the only  
one free parameter (the maximal value of transversal velocity)    
of the measured  $\Omega$ \cite{Omega,Omega1}, $J/\psi$ and $\psi^\prime$ \cite{Jpsi}  
transverse momentum spectra  
in Pb+Pb collisions at 158 GeV$\cdot$A  
that are frozen-out at hadronization temperature $T_H$.  
The experimental  situation with the $\phi$ mesons at SPS   
is, unfortunately,  not clarified yet  because the  results of  
the NA49  \cite{phiA}  and NA50 \cite{phiB}  Collaborations are not in agreement.  
The  analysis of the  transverse momentum spectra of  
$\Omega$ hyperons  \cite{BUG:02C,Suire:02} 
and $\phi$ mesons  \cite{BUG:02C} 
reported   by the STAR Collaboration for  energies  
$\sqrt{s} = 130$ A$\cdot$GeV in Refs. \cite{OMEGA:130} and \cite{phi:130}, respectively,  
  and for   
$\sqrt{s} = 200$ A$\cdot$GeV in Ref. \cite{Suire:02}    
shows that this picture remains valid for RHIC energies as well.  
 
It is easy to find out that for such particles like  
$\phi$, $\Omega$, $J/\psi$ and $\psi^\prime$ 
which  are  weakly interacting   with other hadrons  
the distribution function $\phi_{out}$ 
should coincide with  $\phi_{in}$ 
 
\begin{equation}\label{foone} 
\hspace*{-0.5cm}
\phi_{out} \Bigl|_{\Sigma^*}    =   
\phi_{in}~ (~T_{in}^*, ~\mu_{in}^*, v^*_{in})~~ \Theta(~~ p^1  \partial_1 {\cal F}^*_L )  \,, 
\end{equation}  
where, in contrast to (\ref{thirty}), there  is no incoming component of 
the distribution because the non-interacting particles cannot rescatter  
and, hence, change their velocity. 
Note also that a small modification of the   
incoming part of
$J/\psi$ momentum distribution  due to decay of heavier charmonia in the ``out'' domain
can be safely neglected.
Remarkably,  the cascade initial condition (\ref{foone}) exactly coincides  
with the one used in the TLS model. Therefore,  
the main TLS conclusions \cite{SHUR:01} on the $\phi$ mesons and  $\Omega$ hyperons  
remain unchanged, whereas  for hadrons with large scattering cross-sections the TLS 
conclusions may change significantly.

Omitting the contributions of weakly interacting hadrons  from the components
of the energy-momentum tensor and baryonic  4-current,
one can generalize the boundary conditions
(\ref{twsix}) and (\ref{twseven}) on the  hypersurface $\Sigma^*$ between  
the domains, and formulate the energy-momentum and charge conservation laws 
in terms  of the {\it cut-off}  distribution functions.  
For definiteness we shall consider the first order phase transition between 
QGP and hadronic matter through out the rest of this work. The case of second order  
phase transition can be analyzed similarly.  
Thus, in terms of the local coordinates $(t^*_L (\bar x_L ); \bar x_L )$, 
introduced in Sect. 5, the conservation laws (\ref{twsix}) and (\ref{twseven}) 
can be generalized as follows  ($ \nu = \{0;~1\}$) 
\begin{eqnarray} 
\hspace*{-1.0cm}
&&
\alpha_q 
\sum_{Q = q, \bar{q}, \ldots} 
T^{1 \nu}_{Q \pm} ~(~T_{in}^*, ~Z_Q\cdot\mu_{in}^*, ~ v^*_{in})  + \nonumber \\
\hspace*{-1.0cm}
&& (1 - \alpha_q)  \sum_{H = \pi, K, \ldots} 
T^{1 \nu}_{H \pm} (~T_{in}^*, ~Z_H\cdot
\mu_{in}^*, ~ v^*_{in})  = \nonumber \\
\label{fotwo} 
\hspace*{-1.0cm}
&& 
\sum_{H = \pi, K, \ldots} 
T^{1 \nu}_{H \pm} (T_{out}^\pm, Z_H\cdot\mu_{out}^\pm, v^\pm_{out})
\,, \\  
%
%
\hspace*{-1.0cm}
&& 
\alpha_q 
\sum_{Q = q, \bar{q}, \ldots} 
N^{1 }_{Q \pm} ~(~T_{in}^*, ~Z_Q\cdot\mu_{in}^*, ~ v^*_{in})  + \nonumber \\
\hspace*{-1.0cm}
&& (1 - \alpha_q)  \sum_{H = \pi, K, \ldots} 
N^{1 }_{H \pm} (~T_{in}^*, ~Z_H\cdot\mu_{in}^*, ~ v^*_{in})  = \nonumber \\
\label{fothree} 
\hspace*{-1.0cm}
&& 
\sum_{H = \pi, K, \ldots} 
N^{1 }_{H \pm} (T_{out}^\pm, Z_H\cdot\mu_{out}^\pm, v_{out}^\pm)
\end{eqnarray} 
where $\alpha_q$ is the volume fraction of the QGP in a  mixed phase, 
the $Q$-sums of the energy-momentum tensor and baryonic 4-current components,
denoted as 
%
\begin{eqnarray}
%
&&  
\label{fofour}
 T^{\mu \nu}_{Q \pm}  \equiv
\int \frac{d^3 p }{p^ 0}~ p^\mu p^\nu ~ \phi_{Q} (x, p) ~  
\Theta(\pm~p^\rho \partial_\rho {\cal F}^{*}) \, , \\
%
 && 
\label{fofive}
 N^{\mu }_{Q \pm}  \equiv
\int \frac{d^3 p }{p^ 0}~ p^\mu ~ Z_Q \phi_{Q} (x, p)~   
\Theta(\pm~p^\rho \partial_\rho {\cal F}^{*}) \, , 
\end{eqnarray} 
run over all corresponding  degrees of freedom of QGP. Similarly,
the  $H$-sums run  
over all hadronic degrees of freedom. In Eqs. (\ref{fotwo}) and (\ref{fothree}) 
$Z_Q$ and $Z_H$ denote the baryonic  charge of the corresponding particle species.

Now from Eqs. (\ref{fofour}) and (\ref{fofive})  it is clearly seen that the correct 
hydro+cascade approach  requires the knowledge of a more detailed  information 
on the microscopic properties of QGP than it is  usually provided by the traditional
equations of state.  To proceed further  we, however, shall assume that those 
components are known. The general approach to calculate  the angular and momentum
integrals in Eqs. (\ref{fofour}) and (\ref{fofive}) was developed in Ref. \cite{BUG:99a}
and was  applied to  the massive Boltzmann gas description in \cite{LPC:TN,BUG:99a}.

The important difference between  conservation laws 
(\ref{fotwo}), (\ref{fothree})  and (\ref{twsix}), (\ref{twseven}) is that 
in the ``out'' domain  
the temperature $T_{out}^-$, chemical potential $\mu_{out}^-$ and relative velocity $v_{out}^-$ 
of the incoming to $\Sigma^*$ hadrons should differ 
 from the corresponding quantities $T_{out}^+$, $\mu_{out}^+$ and $v_{out}^+$ of the outgoing 
from $\Sigma^*$ particles, 
and both sets should  differ from the quantities  $T^*_{in}$, $\mu_{in}^*$ and $v^*_{in}$ of the ``in'' domain.
In order to prove this statement it is necessary to 
compare  the number of equations and number of unknowns for the two distinct cases,  
namely, (i) if the initial state  is in  the mixed QGP - HG  phase, and (ii) if  the initial state 
belongs to the  QGP. 

In case (i) there are 10 equations and 10 unknowns:

$\bullet$ The  equations are as follows: 
6 conservation laws from Eqs. (\ref{fotwo}) and (\ref{fothree}); 
 value of the  initial energy;  value of the initial baryonic density; the relation between 
initial temperature $T^*_{in} $ and the baryonic chemical potential $\mu^*_{in}$ taken
at the phase boundary; and the switch off criterion.

$\bullet$ The unknowns are as follows: three temperatures $T^*_{in}$, $T_{out}^-$, $T_{out}^+$;
three chemical potentials $\mu_{in}^*$, $\mu_{out}^-$, $\mu_{out}^+$; three
velocities $v^*_{in}$, $v_{out}^-$, $v_{out}^+$ defined in the rest frame  of  a discontinuity;
and the QGP fraction volume $\alpha_q$.   
Thus, in this case one can find a desired solution of the system of ten transcendental equations,
which is the most general form of the three flux discontinuity introduced 
by Eqs. (\ref{thirty} ) - (\ref{thtwo}).

To  complete the solution of  hydro equations  (\ref{thsix}), (\ref{thseven}), (\ref{thnine})
and (\ref{forty}) one has to find out  the value of velocity  $v^*_{in}$ from the system of
ten transcendental equations discussed above.  Then this velocity
defines an ordinary  differential equation 
$d x^1_L / d t^*_L =  - v^*_{in}$  for the hypersurface $\Sigma^*$ in the rest frame of matter of the ``in'' domain, which has to be solved
simultaneously with the hydro equations.

If initial state belongs to the interior of the QGP phase, case (ii), 
then the usual  hydro solution will be valid till
the system reaches the boundary with the mixed phase, from which the non-trivial 
discontinuity described by Eqs. (\ref{fotwo}) and (\ref{fothree}) will start on.
Now it is clear what are  the distinctive features from  the previously considered case: 
in contrast to case (i)  on the boundary with the mixed phase 
the volume fraction of QGP  is fixed to unit $\alpha_q = 1$;
 the energy and baryonic charge densities are not independent anymore, but
are completely  defined by the temperature and baryonic  chemical potential, which, 
in addition,  are connected by the entropy conservation for the continuous hydro solution
in QGP.

Therefore, in case (ii) there are 9 equations and 9 unknowns, which are as follows:

$\bullet$ The  equations are: 
6 conservation laws from Eqs. (\ref{fotwo}) and (\ref{fothree}); 
temperature dependence of  the baryonic chemical potential $\mu^*_{in} = \mu^*_{in} (T^*_{in} )$   due to the
entropy conservation; the relation connecting  
 temperature $T^*_{in} $ and  baryonic chemical potential $\mu^*_{in}$, since they belong
to the phase boundary; and the switch off criterion.

$\bullet$ On the other hand the unknowns, except for the fixed volume fraction $ \alpha_q = 1$,
 are the same as in case (i). Thus, again the number of unknowns matches the number of
equations, and the procedure to solve  the system of hydro equations 
(\ref{thsix}), (\ref{thseven}), (\ref{thnine}) and (\ref{forty}) simultaneously with the
 boundary conditions (\ref{fotwo}) and (\ref{fothree}) is the  same as in case (i).


Now it is appropriate  to discuss  
the switch off criterion ${\cal F}^{*}(t, \bar{x}) = 0$  in more details.   
By the construction of  the hydro+cascade  approach,   
the cascade treatment should be applied since the very moment, where the hydrodynamics   
starts to lose its applicability:
according to the original assumption
 the hydro equations (\ref{thsix}), (\ref{thseven}), (\ref{thnine}) and (\ref{forty}) work well
inside of the 4-volume surrounded by 
the hypersurface $\Sigma^*$  and in the outer $\varepsilon$-vicinity 
($\varepsilon \rightarrow 0$) of  $\Sigma^*$
 [see also a discussion after Eq. (\ref{theight})],  whereas just   outside of 
this domain  the thermal equilibrium dismantles and, hence, one 
has no right to use the {\it cut-off} equilibrium distributions  interior of 
the  ``out'' domain.   
Consequently,  a  switch off criterion should be formulated  
solely for some  quantity  defined in  the  outer $\varepsilon$-vicinity
of hypersurface $\Sigma^*$, and it has to define the  bounds of applicability of 
thermal equilibration and/or hydrodynamic description.
Note that
in   the BD and TLS models this did not matter because  
both groups kept the cascade  initial conditions as close  as possible to the output  of 
hydro.  However, in the  case of the {\it three flux discontinuity}  on the time-like 
parts of  hypersurface $\Sigma^*$ the proper use 
of the  switch off criterion   plays a decisive role in the  
construction  of the mathematically correct hydro+cascade solution (see also a 
discussion of the freeze-out criterion in Refs. \cite{BUG:96,BUG:99a}).  
It is clear that, in contrast to the BD and TLS formulations,   
the switch off criterion
may generate a very sizable effect while applied to interior of  hadronic phase.
It is so, since even a small difference (just a few MeV)  between 
the temperature $T^*_{in}$, which belongs to the phase transition region, 
and temperatures $T_{out}^-$ and $T_{out}^+$ of the ``out'' domain may lead to a tremendous  flow
of outgoing  hadrons  because of the enormous latent heat of the QGP.

\vspace*{1.cm}

\section{Concluding  Remarks}  

In the preceding sections we have derived  the 
system of relativistic kinetic equations which describe the particle exchange 
between two domains separated by the hypersurface of arbitrary properties.
We showed that the usual Boltzmann equation for the following sum of two distributions 
$\Phi(x,p) \equiv \Theta_{in}~\phi_{in}(x,p) + \Theta_{out}~\phi_{out}(x,p) $
automatically follows from the derived system, but not vice versa. 
Integrating the kinetic equations  we derived the system of the hydro+cascade
equations for a single degree of freedom. Remarkably, 
the conservation laws on the boundary between two domains
conserve the incoming and outgoing components of the energy, momentum and
baryonic charge separately, and, hence, there is twice the number of  conservation laws
on the separating hypersurface compared to the usual relativistic hydrodynamics.
Then  we showed that for a single degree of freedom 
these boundary conditions between domains
can  be satisfied  only by a special superposition of two {\it cut-off} equilibrium
distributions for the ``out'' domain. Since the obtained discontinuity, in contrast to usual shocks defined by
two fluxes, has three irreducible fluxes, hence, it is named a {\it three flux discontinuity}.   
It was also shown that the TLS-like choice of the boundary conditions, in contrast to
expectation of  \cite{SHUR:01}, corresponds to an analog of the compression shock in traditional
hydrodynamics, and, therefore, it cannot be used to model the rarefaction process.

Then we showed that  existence of the {\it time-like shocks} \cite{TIMESHOCKa,TIMESHOCKb},
formally rederived by this formalism,  contradicts, nevertheless,  to the usual assumptions adopted for the one-particle distributions
and, hence, the solution of this problem  requires the analysis of 
higher order distribution functions.
Therefore, in the rest of the paper  we concentrated on 
a detailed  analysis of the discontinuities at the
time-like hypersurfaces, i.e.  the space-like shocks in terms of Refs. \cite{TIMESHOCKa,TIMESHOCKb}.  These results were then generalized to a more realistic
case, namely, if  the mixed QGP-HG phase  is assigned to the ``in'' domain and hadrons exist 
in the ``out'' domain. Such a generalization  also required the exclusion of the hadrons with
the small scattering cross-section (like $\Omega$, $J/\psi$ and $\psi^\prime$ particles) from the boundary conditions between domains. 
As we showed in the preceding section, the presence of  the first order phase transition makes the resulting 
system of transcendental equations more complicated than in the case of a single degree of freedom.

It turns out, that 
a  minimal number of  variables in this discontinuity is either 9 or 10, depending on the
location of the initial state on the phase diagram, and, therefore, on the hadronic  side
the {\it three flux discontinuity}
 should have  two different flows with their own temperatures, chemical potentials and
collective velocities.
The found   solution has a number of unique features in comparison with usual shocks:

\begin{itemize}
\item  this discontinuity may generate  a very strong, explosive like, flow of outgoing particles 
from the  ``in'' domain, first,  because  a huge latent heat of QGP is involved, and,
second, due to an extra  momentum associated with the {\it cut-off} distribution.
Indeed, considering the outgoing component of the distribution 
$\phi_{out} ~ \Theta( p^1  \partial_1 {\cal F}^*_L ) $  for massless pions  in the
frame where this function maximally resembles the non-cut Boltzmann distribution, 
i.e. in the rest-frame of the latter,
one finds  a nonvanishing  collective velocity $v_\pi = \frac{ (1 + v_\sigma) }{2}$.
Here  $v_\sigma \equiv  \frac{ d R_\perp }{d t} ~(|v_\sigma| \le 1$  for time-like parts of $\Sigma^*$) denotes 
 the transversal radius velocity in this frame.
\item the strong explosive flow of outgoing particles is localized at the time-like parts of the 
hypersurface $\Sigma^*$, whereas  at the space-like parts of $\Sigma^*$
there will be a continuous flow. It is even  possible that for some choice of parameters
the space-like boundary may be absent.
\item the particle density  of outgoing pions will  strongly depend  on the 
speed of the transversal radius expansion. Thus,  for  massless pions
the particle density found according to the Eckart definition \cite{GROOT} is
$\rho_\pi = \frac{ \rho_\pi (T^+_{out})  }{4} \sqrt { (1 -  v_\sigma)^3 (3 + v_\sigma) }$, 
i.e. it is smaller for all $v_\sigma > -1 $ than the thermal particle density $\rho_\pi (T^+_{out}) $.
Therefore, the two particle correlations off  the low particle density regions should be
reduced. Since, the situation   $v_\sigma > > -1 $ is typical  for the beginning of 
the transversal expansion \cite{SHUR:01}, then  the main contribution to the transversal pion correlations  
will come from the  later times of expansion. Thus, it is possible that the space-time region which 
defines 
the side and out  pion correlation radii will be essentially more localized both in space and time 
than in traditional hydrodynamic solutions.
\item since there are two fluxes in the ``out'' domain, they will interact with each other.
The resulting distribution should be, of course, found by the cascade simulations, but
it is clear that the fastest  of them will decelerate and the cold one will reheat.
Besides the possibility  to accelerate or decelerate the outgoing transversal  flow more rapidly than in the BD and
TLS models,  the {\it three flux discontinuity} may naturally generate some {\it turbulence} patterns
in the ``out'' domain. 
\end{itemize}
Taking into account all these features alone with the fact that neither the BD nor TLS boundary
conditions have such a strong discontinuity, we conclude that the {\it three flux discontinuity}
opens a principally new possibility not only to resolve the HBT puzzle \cite{GYULASSY:01},
but to study some  new phenomena, like a turbulence pattern, associated with 
a new kind of shock, 
a {\it three flux discontinuity}, in relativistic hydro+cascade approach.

Despite the reasonably good description of the one-particle
spectra of the most abundant hadrons, even such sophisticated model as the TLS one
badly overestimates both
of the transverse radii measured by pion interferometry  like other hydrodynamic models.
This is a strong indication that the hydro part of  all existing hydro+cascade and
hydrodynamic  models requires an essential revision.
How this revision will affect the present BD and  TLS results is unclear at the moment,
but the solution of the HBT puzzle \cite{GYULASSY:01} should serve as a good
test for the correct picture of the space-time evolution during the post-hadronization stage.
The  additional tests for  the
correct  hydro+cascade equations  should be 
the reproduction of  three  recently  established  signals of the deconfinement  phase transition,
i.e.
the pion kink \cite{Horn,Kink} seen at lab energy of $\sim 30$ GeV$\cdot$A , 
 the  $K^+/ \pi^+$ peak  at the same lab energy  \cite{Horn}  
(the strangeness horn)
and the  plateau \cite{Step} in the inverse slope of the $K^\pm$  transverse momentum 
spectra 
at the whole range of the SPS energies (the step in caloric curves)  
 measured by the  NA49 Collaboration  \cite{SPSKaonsA,SPSKaonsB}.
Also it is   necessary to check  other predictions of the Statistical Model of the Early Stage \cite{Horn}, 
namely the anomalies in  the  entropy to energy fluctuations  \cite{Fink} (the `` fink'') and
in 
strangeness to energy fluctuations \cite{Well}  (the ``tooth''), because  both the ``fink'' and  ``tooth''
may be sensitive to 
the turbulence behavior due to  energy dissipation.

Note, however, that the completion of this task requires an additional research of the 
hydro+cascade approach. First, it is necessary to develop further the microscopic models
of the QGP equation of state in order to find out the required by Eqs. (\ref{fotwo}) - (\ref{fofive})
components of the {\it cut-off}  energy-momentum tensor and baryonic 4-current.  
This can be done, for example, within 
the phenomenological extensions \cite{MIG:81,JL:94,DBB:03} of  
the  Hagedorn model. Second, a similar problem for hadrons should be solved as well, otherwise,
as  we  discussed in preceding section,
the {\it switch off} criterion from the hydro to cascade cannot be formulated correctly 
within the hydro+cascade approach.
And, finally,  for practical modeling it is necessary to formulate a mathematical algorithm to  solve simultaneously  the system of hydro+cascade equations (\ref{thsix}) - (\ref{forty})  with the boundary conditions (\ref{fotwo}) and (\ref{fothree})
between the hydro and cascade domains. 
These problems, however,  should be considered elsewhere.

\vspace*{0.4cm}  
  
\noindent  
{\bf  Acknowledgments.}    
The author   
is thankful to     
D.~Blaschke, 
W.~Cassing and L. W. Phair
for valuable comments, and to P. Huovinen and B. R. Schlei for stimulating  discussions.   
Also I appreciate  very interesting discussions on this subject  with L. P.~ Csernai,  E. V. ~Shuryak and D.~ Teaney.
The author  thanks  the Institute for Nuclear Theory at the University of Washington  
for its warm  hospitality and the Department of Energy for the partial support during the  
completion of this work.   
Also the partial  financial support of the  DFG grant No. 436 UKR 17/13/03 is greatly acknowledged.

  
\def\np#1{{Nucl. Phys.} {\bf #1}}  
\def\prl#1{{Phys. Rev. Lett.} {\bf #1}}  
\def\jp#1{{J. of Phys.} {\bf #1}}  
\def\zp#1{{Z. Phys.} {\bf #1}}  
\def\pl#1{{Phys. Lett.} {\bf #1}}  
\def\pr#1{{Phys. Rev.} {\bf #1}}  
\def\hip#1{{Heavy Ion Phys.} {\bf #1}}  
\def\prep#1{{Phys. Rep.} {\bf #1}}  
\def\preprint#1{{\it  Preprint} {\bf #1}}

\end{multicols}  
 
\end{document}